# First Results from the La Silla-QUEST Supernova Survey and the Carnegie Supernova Project

E. S. Walker[1], C. Baltay[1], A. Campillay[2], C. Citrenbaum[1], C. Contreras[2,4], N. Ellman[1], U. Feindt[3], C. González[2], M. L. Graham[7], E. Hadjiyska[1], E.Y. Hsiao[4], K. Krisciunas[5], R. McKinnon[1], K. Ment[1], N. Morrell[2], P. Nugent[6,7], M.M. Phillips[2], D. Rabinowitz[1], S. Rostami[1], J. Serón[2], M. Stritzinger[4], M. Sullivan[8], B.E.Tucker[9]

1 Physics Department, Yale University, 217 Prospect Street, New Haven, CT 06511-8499
2 Carnegie Institution of Washington, Las Campanas Observatory, Colina el Pino, Casilla 601, La Serena, Chile
3 Inst. fur Physik, HU Berlin and Physikalisches Institut, Universitat Bonn, Regina-Pacis-Weg 3, 53113 Bonn, Germany
4 Department of Physics and Astronomy, Aarhus University, NY Munkegade 120, DK-8000 Arahus C, Denmark
5 Physics Department, Texas A&M University, 4242 TAMU, College Station, TX 77843-4242
6 Lawrence Berkeley National Laboratory, Department of Physics, 1 Cyclotron Road, Berkeley, CA 94720
7 University of California at Berkeley, Berkeley, CA 94720
8 Department of Astronomy, University of Southampton, SO1BJ, United Kingdom
9 School of Astronomy and Astrophysics, Australian National University, Acton ACT 2601, Australia

## ABSTRACT

The LaSilla/QUEST Variability Survey (LSQ) and the Carnegie Supernova Project (CSP II) are collaborating to discover and obtain photometric light curves for a large sample of low redshift (z < 0.1) Type Ia supernovae. The supernovae are discovered in the LSQ survey using the 1 m ESO Schmidt telescope at the La Silla Observatory with the 10 square degree QUEST camera. The follow-up photometric observations are carried out using the 1 m Swope telescope and the 2.5 m du Pont telescopes at the Las Campanas Observatory. This paper describes the survey, discusses the methods of analyzing the data and presents the light curves for the first 31 Type Ia supernovae obtained in the survey. The SALT 2.4 supernova light curve fitter was used to analyze the photometric data, and the Hubble diagram for this first sample is presented. The measurement errors for these supernovae averaged 4%, and their intrinsic spread was 14%.

*Subject keywords*: Cosmology, Supernovae

1. INTRODUCTION

Type Ia supernovae (SNe Ia) played a crucial role in the discovery of the acceleration of our Universe (Riess et al. 1998, Perlmutter et al. 1999). The existence of a dominant new component of the energy density of the Universe, referred to as dark energy, was hypothesized to explain this acceleration, but its nature is very poorly understood. Suggested models include Einstein's cosmological constant or some other form of new scalar field, or the apparent acceleration might signal the need for some modification of General Relativity at cosmological scales. Since the discovery of the acceleration of the universe there have been a number of surveys to pursue this issue using the supernova technique, the major ones being the Equation of State Supernova Cosmic Expansion Survey ESSENCE (Wood-Vasey et al. 2007), the Supernova Legacy Survey SNLS (Conley et al 2011, Sullivan et al 2011), the Sloan Digital Sky Survey SDSS ( Campbell et al. 2013), the Hubble Space Telescope Supernova Searches (Riess et al 2007, Suzuki et al 2012), and the PanSTARRS1 Survey (Rest et al 2014). There were other surveys using different techniques, such as Baryon Acoustic Oscillations:  SDSS (Eisenstein et al 2005), Six degree Field Galaxy Survey 6dFGS (Beutler et al 2011), WiggleZ (Blake et al 2011), the Baryon Oscillations Spectroscopic Survey BOSS (Anderson et al 2012), and Weak Lensing:  SDSS (Huff et al 2011), the Canadian-France-Hawaii Telescope Lens Survey CFHTLens (Benjamin et al 2013) and the Deep Lens Survey DLS (Jee et al 2013). A recent fit to obtain the cosmological parameters using the data from all techniques combined and leaving all of the parameters to vary was carried out by Anderson et al (2012) with the following results: in terms of the accepted parameters, $\Omega_m$=0.270±0.012, $\Omega_k$=-0.010±0.005, $\Omega_{DE}$=0.740±0.013, $w_0$=-0.93±0.16, and $w_a$=-1.39±0.96.

Supernovae will continue to play a key role as one of several techniques that will be used in the future experimental efforts to clarify this situation. The use of supernovae as standard candles is based on a differential measurement between the luminosities of nearby (z < 0.1) and more distant supernovae. Ideally one would like similar numbers of nearby and distant supernovae. The high redshift supernova surveys listed above have collected a distant sample (0.1 < z < 1.0) of the order of 1000 supernovae. There has also been a considerable effort to collect samples of nearby supernovae, such as the Center of Astrophysics Survey CfA (Hicken et al. 2009), the Carnegie Supernova Survey CSP (Hamuy et al. 2006), the Lick Observatory Supernova Search LOSS (Ganeshalingam et al. 2013), the Supernova Factory SNf (Aldering et al. 2002), and the Palomar Transient Factory PTF (Maguire et al. 2014). The size of the nearby sample, however, is considerably smaller than the size of the distant sample. In the most recent analysis of supernova data, the number of nearby (0.01 < z < 0.10) supernovae that were of sufficient quality to be included on the Hubble diagram for Betoule et al. (2014) was 123, and for Rest et al (2014), with somewhat different cuts, was 197. This was a heterogeneous sample: small numbers from different surveys from different instruments analyzed and calibrated in different ways. Thus the quality of the final cosmology fits was limited by the systematic errors, as well as the size, of the nearby sample.

There are ambitious plans for future surveys to collect considerably larger numbers of distant supernovae going out to higher redshifts such as the ground based Dark Energy Survey DES (Bernstein et al. 2012), the Large Synoptic Survey Telescope LSST (LSST Science Book, 2009), and space missions, the Wide Field Infrared Survey Telescope WFIRST (Spergel et al. 2015) and EUCLID (Astier et al. 2011). These large data sets will require a much larger sample of nearby supernovae both to anchor the Hubble diagram and to identify sub-classes of Type Ia supernovae that have a smaller intrinsic spread than the whole population. This will serve to significantly improve on the dark energy constraints obtainable by the high redshift samples. Detailed studies of the supernova survey envisioned by WFIRST show that a sample of 800 to 1000 high quality nearby supernovae could reduce the errors on the cosmological parameters obtainable by the survey by as much as a factor of two (Spergel et al. 2015).

There are, fortunately, a number of surveys underway, motivated to collect larger nearby samples including the Supernova Factory SNf (Aldering et al. 2002), the Carnegie Supernova Project CSPII (Phillips et al., in preparation), Sky Mapper (Keller et al. 2007), the Palomar Transient Factory PTF (Maguire et al. 2014), and La Silla/QUEST, LSQ (Baltay et al. 2013). The goals of these surveys are twofold. One is to reduce the systematic errors of the samples by more careful analysis and calibration methods as well as extending the light curves into the infrared, and the other is to increase the size of the sample of nearby supernovae to come closer to the desired numbers. The first goal has been the focus of the CSP I and II surveys. It is the second goal (as well as the first, of course) that motivates the present LSQ-CSP II collaboration.

The purpose of this paper is to describe the LSQ-CSP II survey, the data analysis and calibration techniques, and to present the first 31 Type Ia supernovae from the survey. Section 2 describes the surveys and the instruments, Section 3 describes the datasets, and Section 4 the data analysis methods and results. Section 5 presents a discussion of the results and the future plans of the survey.

2. SURVEYS AND INSTRUMENTS

All of the supernova described in this paper were discovered in the La Silla/QUEST Southern Hemisphere Variability Survey and were classified spectroscopically as SNe Ia by a variety of larger telescopes. The spectra are available on the WISEREP data base (Yaron & Gal-Yam 2012). The supernovae published here were followed photometrically in multiple filter bands using the Swope telescope at the Las Campanas Observatory to construct the light curves covering the period around maximum light.

2.1 *The La Silla/QUEST Survey*

The La Silla/QUEST survey started the Low Redshift Supernova Search in December of 2011 (Baltay et al. 2013). The survey uses the 1 m ESO Schmidt telescope at the La Silla Observatory in Chile with the 10 square-degree QUEST camera (Baltay et al. 2007) located at the prime focus. The camera consists of 112 CCD detectors with 600x2400 pixels each. The pixels are 13 micron square and correspond to 0.87 arc seconds per pixel. With typically 60 s exposures in a broad g + r filter (4000 to 7000 Angstroms) the limiting magnitude is around 21.5. About 1000 square-degrees are scanned on a clear night and each field is imaged at least twice during the night at two-hourly intervals to reduce contamination from Solar System objects that move across the image.  The nightly pattern is repeated with a two-day cadence. The search is blind in that it is sensitive to supernovae regardless of their proximity to or the type of host galaxies. Reference images, consisting of co-added images taken at least two weeks before the discovery images, are appropriately normalized and subtracted from the discovery images in order to isolate candidate transients. These initial candidates are filtered by imposing cuts based on the signal-to-noise ratio, point spread function and other shape parameters in order to remove noise artifacts and stellar transients. Images of each of the remaining candidates are visually inspected, along with plots of their historical light curves measured from previous LSQ survey images, to select candidates for spectroscopic classification. Of **the candidates for which spectra were taken** 84% turned out to be supernovae.

2.2 *Spectroscopy*

The spectroscopy to classify the supernova candidates was carried out using five different telescopes. The spectra taken for this sample of supernovae that had peak brightness before May of 2013 are summarized in Table 1. The scarcity of spectroscopy time was one limiting factor in the survey. When spectroscopy time was available those candidates were chosen for spectroscopy that were deemed to be the youngest after supernova explosion. The spectra were classified using the SNID (Blondin and Tonry 2007) and the GELATO (Harutyunyan et al 2008) supernova classification programs. For good signal to noise spectra the two were found to give consistent results. All of these supernova classifications were announced promptly in the online reports of the Astronomer's Telegram (ATELs).

**Table 1.**
Spectroscopic selection of Type Ia Supernovae that peaked before May 2013 [*]

| Source | Telescope | Spectrometer | No. of SNe |
|---|---|---|---|
| PESSTO | 3.5 m NTT | EFOSC-II | 12 |
| CSP II | 2.5 m du Pont | WFCCD | 10 |
| SNfactory | 2.2 m UHT | SNIFS | 8 |
| LCOGT | 2.0 m Faulkes | FLOYDS | 2 |
| PTF/CalTech | 5.0 m Palomar | DBSP | 1 |

*The acronyms used in this table are as follows: PESSTO-Public ESO Spectroscopic Survey for Transient Objects. CSPII- the second Carnegie

Supernova Survey. SNfactory-Supernova Factory. LCOGT-Las CumbrasOptical Global Telescopes. PTF-Palomar Transient Factory. NTT-New Technology Telescope.UHT-University of Hawaii Telescope. EFOSC-II- ESO Faint Object Spectrograph and Camera. WFCCD- wide Field CCD Camera. SNIFS- Supernova Integral Field Spectrometer. FLOYDS- Faulkes Low Resolution Spectrograph. DBSP- Double Spectrograph on the 200" Telescope.

### 2.3 *The Swope Telescope*

The 1 m Swope telescope at the Las Campanas Observatory was used to follow the SNe Ia with optical imaging to obtain their light curves. The SITe3 CCD camera, a 2048 x 3150 CCD array with 15-micron pixels that correspond to a plate scale of 0.435 arcsec/pixel, was used for these observations. To speed up the readout time, only the central 1200 x 1200 pixels were read out. Exposure times were typically 5 to 10 minutes near peak, longer for fainter supernovae and for supernovae long past peak, with nominally each of six filters, the *B, V* (Bessel 1990), and the SDSS (Fukugita et al. 1996) *u, g, r,* and *i*. This telescope and instrument have been calibrated with great care in the course of previous CSP campaigns (Hamuy et al. 2006, Contreras et al. 2010, Stritzinger et al. 2011). The extinction coefficients and the color terms have been carefully measured and shown to be extremely stable over many years. These color terms and extinction coefficients will be updated (Krisciunas et al. 2015). The **better then 1%** calibration of this instrument and the typically 1 arcsec seeing at the Las Campanas site, is playing an important role in obtaining light curves of the highest quality.

### 2.4 *The du Pont Telescope*

The 2.5 meter du Pont telescope, also located at the Las Campanas Observatory in Chile, has been used both to take spectra to classify supernova candidates and to take the final template images of the host galaxy after the supernova has faded to allow the subtraction of the host galaxy light from the measured light curve. For the spectroscopy the WFCCD spectrometer was used with a wavelength range of 3800 to 9200 Angstroms. For template imaging the detector used is a SITE2 2048 x 2048 CCD with a plate scale of 0.259 arcsec/pixel. The du Pont is preferred to the Swope to take the generally fainter host galaxy images due to its larger size and better seeing. For these template exposures the same physical filters were used as were used for the photometric observations with the Swope telescope, facilitating good subtractions. The du Pont telescope was also used with the RetroCam camera to take infrared images of a subset of these supernovae in the *Y, J,* and *H* bands for another part of the CSP II low redshift supernova project.

3. DATA SET

In the first two years of the LSQ survey, 399 supernova candidates had a good quality spectrum taken. Of these, 336 or 84%, turned out to be classified as supernovae, 237 SNe Ia's, 25 SNe Ib,c's, and 74 SNe II's. Photometric follow up of a total of 107 of the supernovae were initiated on the Swope telescope, some even before spectroscopic classification. Of the spectroscopically confirmed SNe Ia's 55 had light curves that started near or before peak brightness with six or more observations in multiple filters. Of these 31 had reached peak brightness before May of 2013 and were thus faded sufficiently so that the galaxy template image could be taken. This first sample of 31 supernovae forms the basis of this paper.

The survey was unbiased with respect to the presence or absence of a host galaxy since it did not target galaxies but was a blind search depending only on the increased brightness of a point like source. The completeness of the supernova discoveris below a redshift of 0.1 was measured in the early tuning of the search selection criteria, **by inserting simulated sources in the images, to be** on average 85%. The most significant loss was of supernovae near the center of bright host galaxies. The selection of candidates for spectroscopic follow up was based on the estimated age of the candidate and its brightness and was thus not biased with respect to the nature (or absence) of a host galaxy.

The heliocentric redshift of each supernova was obtained by one of several methods. For 20 of the supernovae the redshifts were obtained from the known redshift of the host galaxy from the NED catalog. These have a redshift error $\delta z = 0.0001$, given in column 5 of Table 3. For another 4 supernovae the redshift was obtained from [OII] or Hα lines from the host galaxy in the spectra taken to type the supernova. These have an estimated z error of 0.001. Finally, for 7 supernovae the redshift was obtained from the spectral features of the supernovae from their typing spectra. These have an estimated z error of 0.005. The distribution in these redshifts for this sample is shown in Figure 1. The heliocentric redshifts were used in the SALT2.4 fitting of the supernovae. Using the RA and dec of each supernova their CMB redshifts $z_{CMB}$ were calculated. The CMB redshifts were used to place the supernovae on the Hubble diagram.

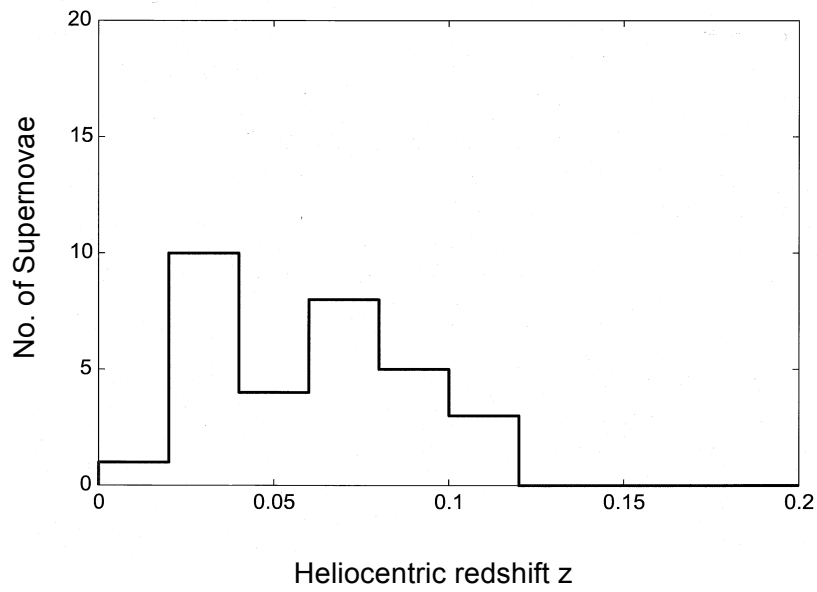

**Figure 1.** Distribution in the heliocentric redshifts of the Type 1a sample.

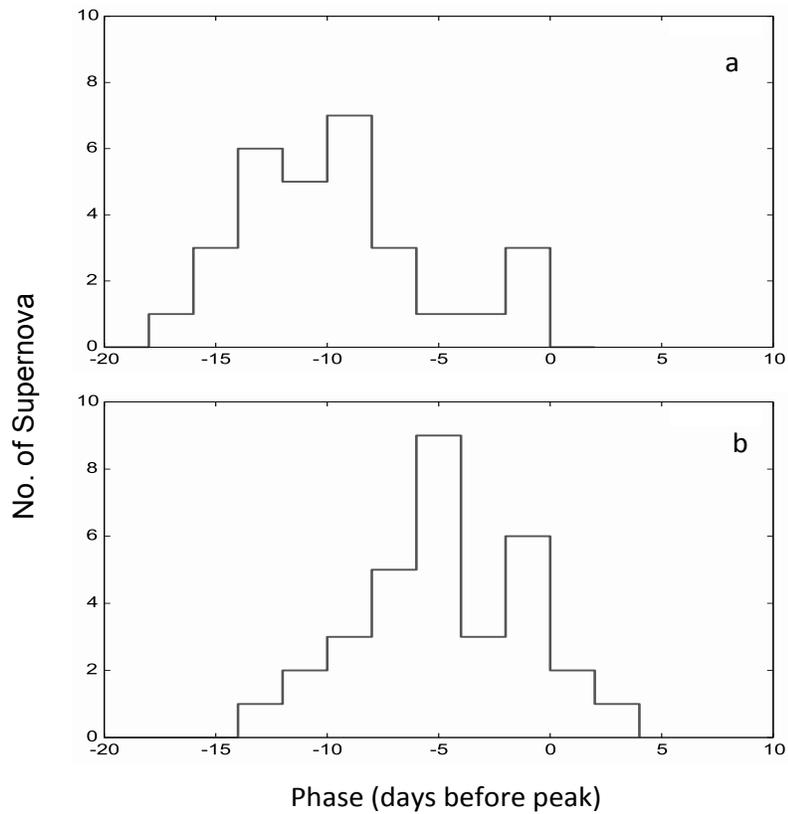

**Figure 2.** Phase (days before *B* band peak brightness) at a) supernova discovery and b) at first point in the *B* filter on the follow-up light curve measured with the Swope telescope.

The phase (days past peak brightness in the *B* band) at the supernova discovery and the phase at the first point on the follow-up light curve measured with the Swope telescope are shown in Fig. 2. There is a delay time between the supernova discovery and the first point on the Swope light curve. An effort will be made to reduce this delay time as much as possible in the future of this survey.

## 4. DATA ANALYSIS AND RESULTS

The Swope lightcurve data are reduced following the procedures developed by the Carnegie Supernova Project as described in Hamuy et al. (2006) and Contreras et al. (2010).

### 4.1. Preprocessing, Photometry, and Calibration

As the first step images are bias-subtracted and flat-fielded, after which an astrometric solution is applied. Linearity and exposure time variation corrections are also applied. PSF instrumental magnitudes ($m_{inst}$) are then measured for a number of selected field stars (local sequence stars) common to all the images for a given target. We use an IRAF[1]-based pipeline developed by the Carnegie Supernova Project. These local-sequence magnitudes are calibrated using catalog standards (by Landolt, 1992, for the *B* and *V* filters and by Smith et al., 2002, for *u, g, r, i*) on at least three photometric nights to obtain measurements of the zero points. Photometric nights are selected by determining the zero points for all of the standard stars during the night and requiring that the scatter in these zero points be less then 0.03 mags for the *B, V, g, r, i* bands and less then 0.08 mags for the *u* band. For this calibration the Landolt and Smith standards are converted to the natural system of the Swope telescope. This means that we do not transform our observations to the systems of the standard catalogs, which would require the use of S-corrections (Stritzinger et al. 2002). Instead, we transform the catalog standards to the Swope system using stable color terms that have been well characterized from years of extensive observations of stellar standards with the Swope telescope (see equations 1-6 of Contreras et al. 2010). For a given band pass, the transformation from $m_{inst}$ to the Swope natural magnitude system ($m_{nat}$) is then given by

$$m_{i,nat} = m_{i,inst} - k_j X_i + zpt_j,$$

where $zpt_j$ is the zero point measured for given photometric night, $X_i$ is the air mass of that field star, and $k_j$ is the air mass extinction coefficient for band pass j. We use extinction coefficients that have been measured for the Swope telescope site and found to be remarkably stable over the years (see Fig. 3 of Contreras et al., 2010).

1. http://iraf.noao.edu

*4.2 Galaxy Background Light Subtraction*

Before measuring supernova magnitudes, we remove the host-galaxy light by subtracting a host galaxy template image taken after the supernova has faded to a negligible level (>300 days after maximum light). These template images were deeper and taken with seeing conditions that match or exceed the seeing conditions for the light curve images, and were typically taken at the larger du Pont telescope, and occasionally with the Swope telescope, both at the Las Campanas Observatory in Chile. In either case the same physical filters are used as for the supernova light curve exposures. Two algorithms were used to carry out these galaxy subtractions. One was a fast automated procedure using the HOTPANTS image subtraction program (Becker et al. 2004). All supernovae were run through this fast method. The subtracted images were examined to assess how well the subtraction worked. About two thirds of the supernovae were judged to be successfully subtracted by this method. The remaining third were then processed by a slow manually guided subtraction procedure to obtain satisfactory subtractions.

*4.3 Supernova Light Curves*

Using the procedure described above, light curves for each of the supernovae in the sample, using point-spread function fitting for aperture-independent photometry, were obtained in each of the filters in the Swope natural system. The light curves for supernova LSQ13ry are shown in Figure 3. The light curves for the remaining 30 supernovae in this sample are shown in Figure A1 in the Appendix. The lines on these Figures are the SALT2.4 fits. On the whole ( but see comment in the last paragraph of Section 4.4) the light curves are well sampled and well fitted by the SALT2.4 templates (outlier points are ignored in these fits). All of the light curves in the sample are presented in numerical form in Table 2 (the first ten lines are in the printed version, the remainder are available electronically).

**Table 2.**
Supernova Photometric Observations*

| Object | Filter | JD | $m_{nat}$ | $\delta m$ |
|---|---|---|---|---|
| LSQ11bk | B | 2455911.68227 | 17.244 | 0.009 |
| LSQ11bk | B | 2455912.69442 | 17.260 | 0.010 |
| LSQ11bk | B | 2455913.70404 | 17.269 | 0.011 |
| LSQ11bk | B | 2455914.71416 | 17.307 | 0.011 |
| LSQ11bk | B | 2455915.64831 | 17.321 | 0.008 |
| LSQ11bk | B | 2455916.66398 | 17.368 | 0.007 |
| LSQ11bk | B | 2455917.66846 | 17.410 | 0.007 |
| LSQ11bk | B | 2455918.63413 | 17.448 | 0.009 |
| LSQ11bk | B | 2455922.65933 | 17.680 | 0.012 |
| LSQ11bk | B | 2455923.61765 | 17.745 | 0.011 |

\* The third column gives the Julian date of the start of each exposure (to approximately the nearest second). The fourth column of this Table gives the calibrated magnitudes in the Swope natural system, and the fifth column gives the error on the magnitudes. The quoted error is the combination of the measurement error on the magnitude and the statistical error which incorporates errors on the mean zero point and the extinction coefficient.

*4.4 Supernova Template Fitting using SALT2.4.*

We use the SALT2.4 supernova template fitting program (Guy et al.. 2007, Guy et al.. 2009, Betoule et al. 2014) to fit the light curves to obtain the best estimate of the rest frame *B* band peak magnitude for each supernova, the width of its light curve, and its color. SALT2.4 has a large collection of supernova spectra at various redshifts and phases from previous surveys. It uses these templates to fit the light curves in several filter bands and does the K corrections to convert the filter bands to the rest frame of the supernova. SALT2.4 also has the effective transmission curves of the filters used at several telescopes, including the Swope, and thus is able to fit the light curves in the Swope natural system. The inputs to SALT2.4 are the light curves in several filters, the heliocentric redshift, and the Milky Way extinction for each supernova. The Milky Way extinctions are taken from the dust maps of Schlegel et al. (1998) with the Schlafly et al. (2011) recalibration. The output of SALT2.4 is the peak magnitude $m_B$* in the rest frame B band, the stretch factor $x_1$, the color c, and the time of the peak magnitude of the supernova. The color c equals the excess B-V color over the natural B-V of a typical SNe Ia in the SALT synthetic spectrum.

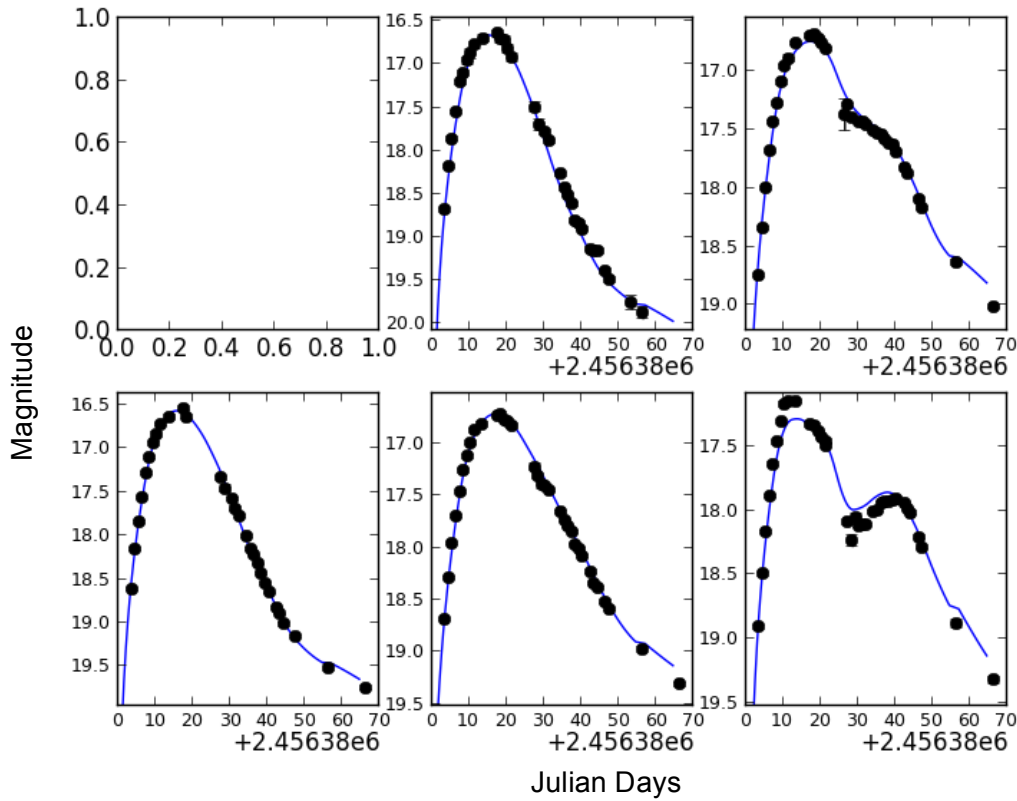

**Figure 3**. Light curves for supernova LSQ13ry in filters *u, B, r* (top row) and *g, V, i* (bottom row). In some cases. measurements were not made in all filters. Missing filter observations are left blank**.** The curves are the SALT2.4 fits.

The values of these output parameters for our sample are summarized in Table 3. The distribution in the stretch parameter $x_1$ from SALT2.4 is given in Figure 4 and the distribution in the color parameter c is given in Figure 5. Superimposed on these Figures are the $x_1$ and c distributions from other supernova samples, some at low redshift (0.01<z<0.1) and some at higher redshifts. **The level of agreement between these samples is discussed in Section 5.1 below.**

The SALT2.4 fits to the light curves are shown as the lines on Figures 3 and A1. A close inspection shows that in a few cases, such as LSQ11ot for example, the fit to the i band is not very good. Since the B and V light curves in these cases are well sampled, removing the i band from the SALT2.4 fit makes a negligible difference, less then a third of a standard deviation in the case of LSQ11ot.

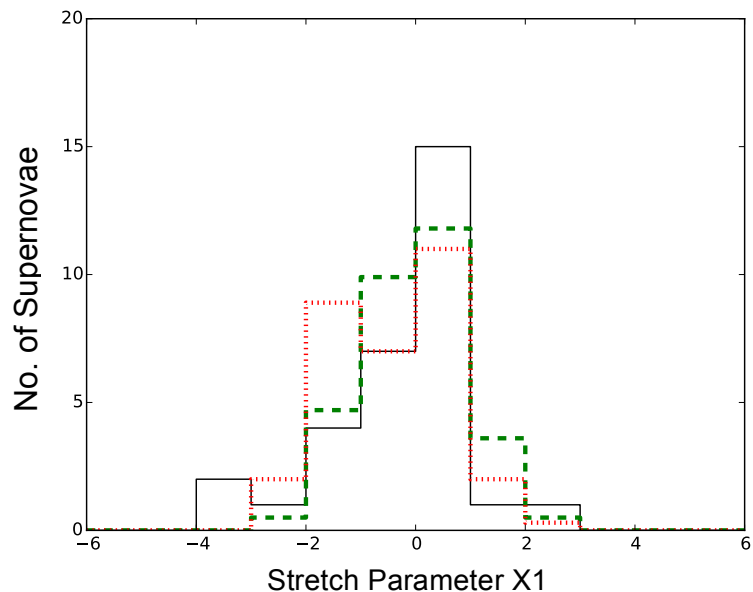

**Figure 4**. Distribution in the stretch parameter $x_1$ from SALT2.4 for this data set (the solid black line). For comparison superimposed are the $x_1$

distributions from PanSTARRS (dashed green line) and the collected low redshift data (red dotted line) from Rest et al (2014). For comparison all three are normalized to the same area.

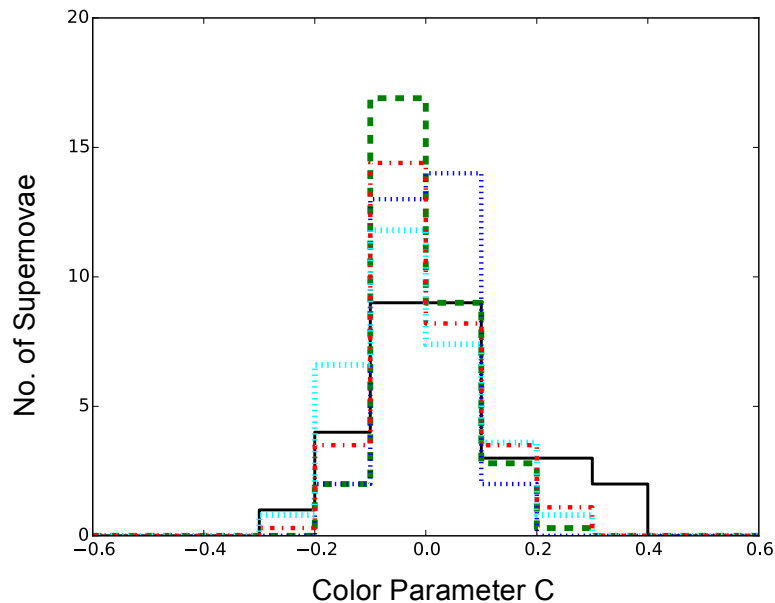

**Figure 5**. Distribution in the color parameter c from SALT2.4 for the present data (solid black line). For comparison are superimposed the c distributions from SNLS (green dashed line) and SDSS (blue dotted line) from Conley et al (2011), and PanSTARRS (blue-green dotted line) and the collected low redshift sample (red x.x. line) from Rest et al (2014). All of the distributions are normalized to the same area for comparison.

*4.5 The Distance Moduli*

The distance modulus is defined as the difference between the apparent and the absolute magnitude of an object. Since the absolute magnitude is defined at a distance of 10 pc from the object, the distance modulus is given by

$\mu = 5.0 \log_{10} (d_L/10pc)$,

where $d_L$ is the luminosity distance (in units of pc) which depends on the cosmological parameters as well as the Hubble constant $H_0$. In the comparison of the distance moduli of our sample with the Hubble curve, i.e. the expected distance modulus $\mu_H$ versus z curve (the curve on Figure 8 below) we use the cosmological parameters from the latest compilation of all of the presently available data (Andersen et al 2012) which are $\Omega_m$ = 0.270±0.012, $\Omega_{DE}$ = 0.740±0.013, $\Omega_k$ = -0.010±0.005, and $w_0$ = -0.93±0.16, and wa=-1.39±0.96, although at the low redshifts of our sample the dependence on these parameters is negligible. For the Hubble constant we use $H_0$ = 70.8±1.4 km/sec/Mpc.

The distance modulus μ of each supernova is determined using the equation

$$\mu = m_B^* - M + \alpha x_1 - \beta c,$$

where $m_B^*$, $x_1$ and c are the SALT2.4 output values as defined above. M is the absolute B magnitude of SNe Ia, and α and β are the stretch and the color correction coefficients that are determined in a fit to the Hubble curve (μ vs z) as described below. The determination of M also depends on $H_0$; thus the value of M we quote below is appropriate for the value of $H_0$ we use in calculating the distance moduli. The values of α and β vary somewhat depending on the choice of the width and color of a "typical" SNe Ia with respect to which the stretch factor $x_1$ and the color c are calculated by SALT2.4.

We evaluate M, α, and β in the following way. Using some starting values of these parameters we calculate the distance modulus $\mu_i$ for each supernova and plot it versus its CMB redshift, i.e. construct the Hubble diagram. We then vary the parameters M, α, and β to minimize the $X^2$ of the points around the Hubble curve. We define the $X^2$ as

$$\chi^2 = \sum_i \frac{(\mu_i - \mu_H)^2}{\sigma_i^2}$$

where the sum is over the number of supernova with distance moduli $\mu_i$, and $\mu_H$ is the distance modulus expected at the redshift of the supernova as discussed above. The $\sigma_i$ are the errors on the distance modulus of supernova i, and they have three ingredients, $\sigma_i^2 = \sigma_{meas}^2 + \sigma_{sys}^2 + \sigma_{int}^2$, where $\sigma_{meas}$ is the measurement error on the corrected B band magnitude, $\sigma_{sys}$ is a systematic error, and $\sigma_{int}$ is the intrinsic spread in the supernova magnitudes. The error due to the uncertainties on the redshifts is negligible, as well as errors due to peculiar velocities for supernovae with redshifts above 0.01. In analyses to measure cosmological parameters (see for example Conley et al 2011, Betoule et al 2014, and Rest et al 2014), supernovae below a redshift of 0.01 are usually eliminated to ensure that the supernovae are in the Hubble flow. None of our sample is below a redshift of 0.01.

The measurement error on each supernova is calculated using the 3x3

correlated error matrix $C_{ij}$ in the variables $m_B^*$, $x_1$ and $c$ which is an output from the SALT2.4 fit. The measurement error on the distance modulus is then taken to be

$$\sigma_{meas}^2 = \mathbf{V}^T \mathbf{C} \mathbf{V},$$

with the vector $\mathbf{V}^T$=(1, α, -β). The distribution in $\sigma_{meas}$ is shown in Figure 6. The average measurement error is slightly over 4% for this sample.

The systematic error $\sigma_{sys}$, includes the systematic errors in the flux calibrations (partly due to the systematic errors in the luminosities of the standard stars used in the calibrations), and systematic errors introduced by the galaxy subtractions, which we believe to be the dominant part. Based on the quality of the subtracted images and the consistency of the result of different subtraction algorithms we estimate the systematic error to be $\sigma_{sys}$ = 0.03 mags. The limited analysis discussed here to estimate the values of M, α, and β is not very sensitive to the value of the systematic errors.

To estimate M, α, and β, fits of the distance modulus of each supernova in this sample to the Hubble curve, (μ vs z) are carried out in an iterative fashion. In the first fit all of the 31 supernovae in the sample are used and the intrinsic spread is fixed at an arbitrary starting value of $\sigma_{int}$ = 0.17 (the exact starting value of $\sigma_{int}$ is unimportant since it will be varied later to get its best fit value). The values of M, α, and β are varied to minimize the $X^2$ of the residuals with respect to the Hubble curve. The number of standard deviations that the distance modulus of each supernova is from the Hubble curve is calculated. The distribution in the absolute value of these standard deviations is shown in Figure 7. The reduced $X^2$ for this fit (the $X^2$ divided by the number of degrees of freedom) was 1.7. Three of the supernovae (LSQ11ot, LSQ12gxj and LSQ13vy) were more then 3 standard deviations from the curve (see Fig 7). Statistically we expect one in a thousand entries to be beyond 3 standard deviations so we consider these three supernovae to be outliers. When these three outliers are removed from the fit, the value of the intrinsic spread $\sigma_{int}$ is varied to make the value of the reduced $X^2$ equal to 1. In this final fit the errors due to the uncertainties on M, α, and β (listed in Table 4) are added in quadrature. This procedure yields the best value of the intrinsic spread $\sigma_{int}$ = 0.14.

All three of the outliers, LSQ11ot, LSQ12gxj and LSQ13vy, are sub-luminous by about 0.6 magnitudes with respect to the rest of the sample. They have significant Na I D absorption lines in their spectra which can be interpreted as a sign of absorption along their line of sight (see for example Poznanski et al. 2012) although Na I D absorbtion may not be a reliable indicator of absorbtion (Phillips et al 2013). LSQ11ot and LSQ12gxj are the most reddened in the sample, suggesting an imperfection in the color-magnitude relation. In any case we are not throwing these three supernovae away. We are merely not using them in the fit to estimate M, α, β and the intrinsic spread for this sample. They remain in Table 3 and their light curves in the Appendix to be included or not as seen fit in any future analysis for the cosmological parameters

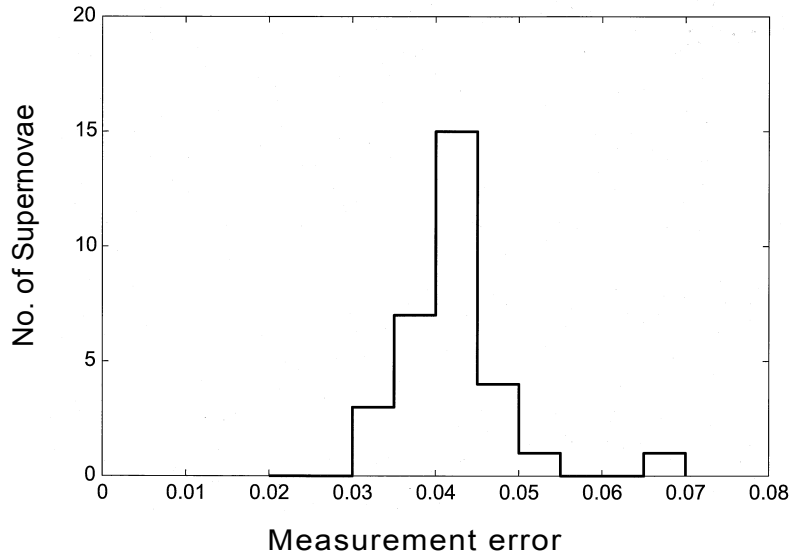

**Figure 6**. Distributions in the measurement errors on the supernova distance moduli.

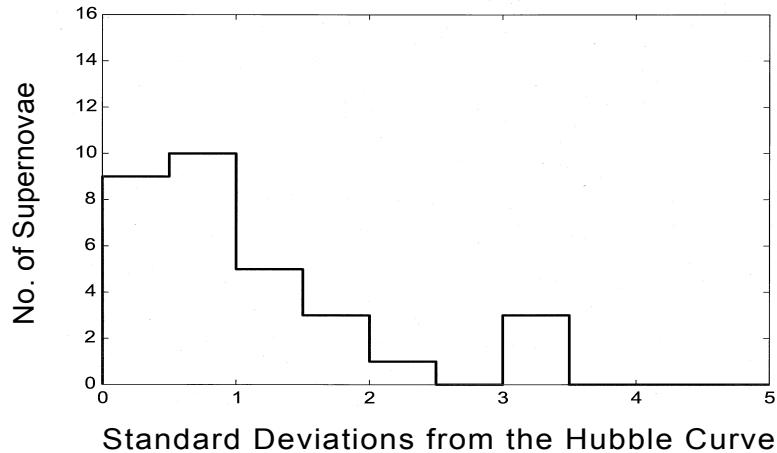

**Figure 7**. Distribution of the absolute value of the number of standard deviations of the distance moduli for each supernova from the expected distance modulus at its redshift with the cosmological parameters given in Section 4.5 (which for short we call the Hubble Curve).

*4.6 The Hubble Diagram*

The results of the $X^2$ fit to the Hubble curve described in Section 4.5 are given in Table 4. With these parameters the distance moduli of the supernovae are calculated and plotted versus their CMB redshift, producing the Hubble diagram for this sample, shown in Figure 8 below. The RMS spread of the points around the Hubble curve is 15.6 %. The error bars in this figure include the measurement errors and the intrinsic spread of the supernovae added in quadrature, combined with the errors due to the uncertainty of the parameters M, α and β.

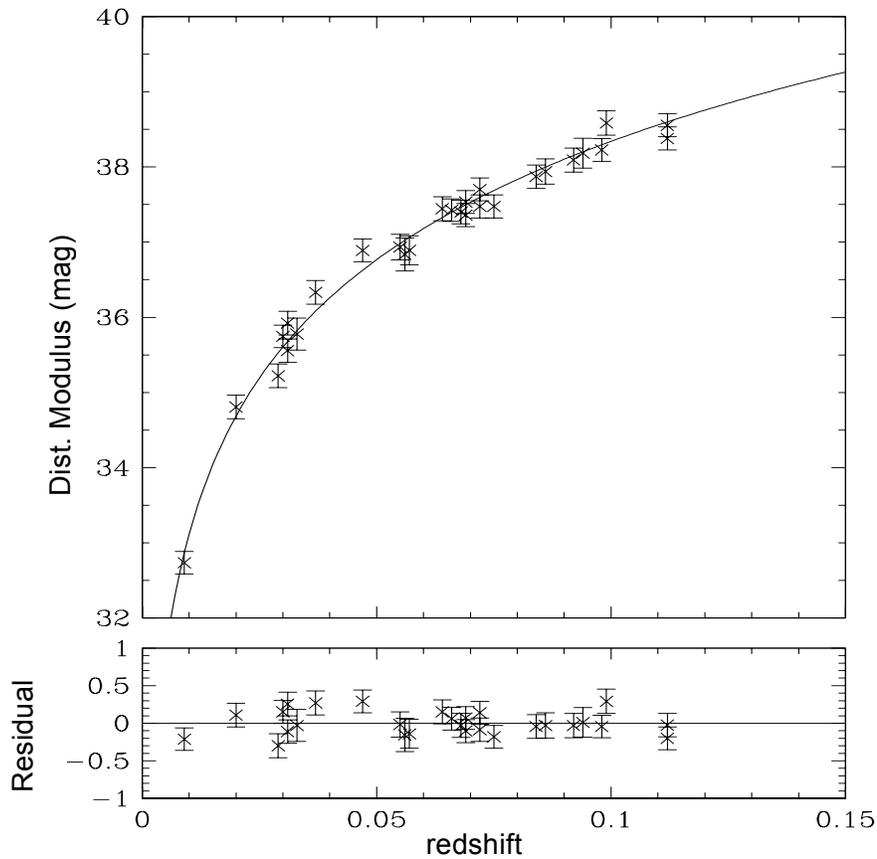

**Figure 8**. Hubble Diagram for the supernovae in this sample. The solid line in this figure is the expected distance modulus versus redshift curve with the cosmological parameters given in Section 4.5.

**Table 3**.

Parameters for the LSQ-CSP supernova sample*

| Object | RA | Dec | $z_h$ | $\delta z$ | $z_{CMB}$ | $m_B^*$ | $\delta m_B^*$ | c | $\delta c$ | $x_1$ | $\delta x_1$ | E(B-V) |
|---|---|---|---|---|---|---|---|---|---|---|---|---|
| LSQ11bk | 04:20:44.25 | -08:35:55.75 | 0.037 | 0.005 | 0.037 | 16.901 | 0.018 | -0.097 | 0.017 | 1.086 | 0.040 | 0.100 |
| LSQ11ot | 05:15:48.34 | 06:46:39.36 | 0.027 | 0.0001 | 0.027 | 17.822 | 0.018 | 0.293 | 0.018 | -0.073 | 0.036 | 0.167 |
| LSQ11pn | 05:16:41.54 | 06:29:29.40 | 0.033 | 0.0001 | 0.033 | 17.413 | 0.019 | 0.136 | 0.019 | -3.016 | 0.073 | 0.152 |
| LSQ12agq | 10:17:41.67 | -07:24:54.45 | 0.064 | 0.0001 | 0.066 | 18.494 | 0.022 | 0.078 | 0.020 | 0.267 | 0.071 | 0.038 |
| LSQ12aor | 10:55:17.64 | -14:18:01.38 | 0.093 | 0.0001 | 0.094 | 19.447 | 0.021 | -0.002 | 0.021 | -2.528 | 0.106 | 0.042 |
| LSQ12bld | 13:42:44.03 | 08:05:33.74 | 0.083 | 0.0001 | 0.084 | 19.022 | 0.020 | 0.048 | 0.020 | -0.847 | 0.070 | 0.023 |
| LSQ12blp | 13:36:05.59 | -11:37:16.87 | 0.074 | 0.0001 | 0.075 | 18.333 | 0.030 | -0.054 | 0.025 | -0.357 | 0.093 | 0.049 |
| LSQ12btn | 09:21:30.47 | -09:41:29.86 | 0.054 | 0.0001 | 0.055 | 18.242 | 0.019 | 0.075 | 0.017 | -1.578 | 0.054 | 0.033 |
| LSQ12ca | 05:31:03.62 | -19:47:59.28 | 0.098 | 0.0001 | 0.098 | 19.122 | 0.021 | -0.016 | 0.018 | -0.010 | 0.160 | 0.043 |
| LSQ12cdl | 12:53:39.96 | -18:30:26.16 | 0.111 | 0.001 | 0.112 | 19.171 | 0.022 | -0.108 | 0.021 | 0.554 | 0.267 | 0.050 |
| LSQ12fuk | 04:58:15.88 | -16:17:58.03 | 0.020 | 0.001 | 0.020 | 15.842 | 0.029 | 0.102 | 0.025 | 0.923 | 0.042 | 0.073 |
| LSQ12fvl | 05:00:50.04 | -38:39:11.51 | 0.056 | 0.0001 | 0.056 | 18.668 | 0.027 | 0.214 | 0.026 | -3.177 | 0.133 | 0.018 |
| LSQ12fxd | 05:22:17.02 | -25:35:47.01 | 0.031 | 0.0001 | 0.031 | 16.315 | 0.024 | -0.024 | 0.020 | 0.896 | 0.073 | 0.022 |
| LSQ12gdj | 23:54:43.32 | -25:40:34.09 | 0.030 | 0.0001 | 0.029 | 15.831 | 0.026 | -0.085 | 0.023 | 0.980 | 0.028 | 0.020 |
| LSQ12gef | 01:40:33.70 | 18:30:36.38 | 0.065 | 0.005 | 0.064 | 18.975 | 0.027 | 0.327 | 0.025 | 0.976 | 0.207 | 0.053 |
| LSQ12gln | 05:22:59.41 | -33:27:51.32 | 0.112 | 0.005 | 0.112 | 18.962 | 0.024 | -0.123 | 0.022 | 0.552 | 0.121 | 0.020 |
| LSQ12gpw | 03:12:58.24 | -11:42:40.13 | 0.058 | 0.005 | 0.057 | 17.432 | 0.026 | -0.032 | 0.025 | 2.386 | 0.240 | 0.063 |
| LSQ12gxj | 02:52:57.38 | 01:36:24.25 | 0.036 | 0.0001 | 0.035 | 18.087 | 0.027 | 0.262 | 0.024 | 0.896 | 0.048 | 0.059 |
| LSQ12gyc | 02:45:50.07 | -17:55:45.74 | 0.093 | 0.001 | 0.092 | 18.768 | 0.025 | -0.058 | 0.024 | 0.945 | 0.251 | 0.021 |
| LSQ12gzm | 02:40:43.61 | -34:44:25.87 | 0.100 | 0.0001 | 0.099 | 19.514 | 0.027 | 0.021 | 0.029 | 0.356 | 0.344 | 0.021 |
| LSQ12hjm | 03:10:28.72 | -16:29:37.08 | 0.070 | 0.005 | 0.069 | 18.266 | 0.025 | -0.111 | 0.023 | -0.358 | 0.069 | 0.025 |
| LSQ12hno | 03:42:43.25 | -02:40:09.76 | 0.048 | 0.0001 | 0.047 | 17.920 | 0.027 | 0.056 | 0.024 | 0.187 | 0.155 | 0.106 |
| LSQ12hvj | 11:07:38.62 | -29:42:40.96 | 0.071 | 0.0001 | 0.072 | 18.403 | 0.025 | 0.043 | 0.023 | 0.741 | 0.088 | 0.047 |
| LSQ12hxx | 03:19:44.23 | -27:00:25.68 | 0.069 | 0.0001 | 0.069 | 18.193 | 0.025 | -0.012 | 0.023 | 0.541 | 0.113 | 0.016 |
| LSQ12hzj | 09:59:12.43 | -09:00:08.25 | 0.029 | 0.001 | 0.030 | 16.481 | 0.027 | -0.112 | 0.023 | -0.406 | 0.038 | 0.058 |
| LSQ12hzs | 04:01:53.21 | -26:39:50.15 | 0.072 | 0.0001 | 0.072 | 18.810 | 0.025 | 0.094 | 0.023 | 0.236 | 0.085 | 0.023 |
| LSQ13abo | 14:59:21.20 | -17:09:09.34 | 0.067 | 0.0001 | 0.068 | 18.835 | 0.029 | 0.170 | 0.025 | -0.922 | 0.135 | 0.093 |
| LSQ13aiz | 13:15:14.81 | -17:57:55.65 | 0.009 | 0.0001 | 0.009 | 13.770 | 0.018 | 0.040 | 0.017 | -0.117 | 0.044 | 0.082 |
| LSQ13pf | 13:48:14.35 | -11:38:38.58 | 0.085 | 0.005 | 0.086 | 19.758 | 0.027 | 0.316 | 0.026 | -1.380 | 0.246 | 0.057 |
| LSQ13ry | 10:32:48.00 | 04:11:51.75 | 0.030 | 0.0001 | 0.031 | 16.330 | 0.021 | -0.296 | 0.021 | -1.025 | 0.032 | 0.042 |
| LSQ13vy | 16:06:55.85 | 03:00:15.23 | 0.032 | 0.005 | 0.032 | 17.715 | 0.018 | 0.090 | 0.017 | -1.164 | 0.035 | 0.069 |

* Columns 4 and 5 give the heliocentric redshift with its error, column 6 is the redshift with respect to theCMB. Columns 7 to 12 give the values output by SALT2.4: the restframe blue band magnitude and its error, the supernova color and its error, and the stretch factor and its error. The last column gives the Milky Way extinction color parameter.

**Table 4**
Results of the $X^2$ fit to the Hubble Diagram

| Quantity | Best Fit Value |
|---|---|
| M | 19.07±0.03 |
| α | 0.13±0.05 |
| β | 2.23±0.30 |
| $\sigma_{intrinsic}$ | 0.14 |

5. DISCUSSION AND FUTURE PLANS

We have presented in this paper the first data release of 31 SNe Ia observed and analyzed by the LSQ-CSPII collaboration. The methodology of the analysis of the data is described in some detail and the resulting light curves are presented.

*5.1 The stretch and color parameters*

Examining our resulting distributions in the SALT2.4 output stretch parameter $x_1$ and the color parameter c, we find that the ranges in $x_1$ and c for this sample are quite consistent with the ranges obtained in other larger samples of supernovae. In the $x_1$ distribution in Figure 4 we superimpose the $x_1$ distributions from Figure 12 of the most recent analysis of the cosmological parameters with the PanSTARRS supernova sample (Rest et al 2014) where they collect all of the available low-z (0.01< z <1.0) sample, 197 supernovae, and also show the distribution for the 113 higher redshift supernovae from the PanSTARRS sample. The comparison curves are normalized to the same area as the LSQ-CSPII sample. The distribution in the color parameter c from the Rest et al 2014 paper for the low-z sample and the distribution for the PanSTARRS sample are shown superimposed on Figure 5. In addition we show the c distributions from the SNLS sample of 242 higher redshift supernovae and the SDSS sample of 93 supernovae from Figure 1 of Conley et al (2011) superimposed on Figure 5 above. The comparison curves are normalized to the area of the LSQ-CSPII sample.

**A Kolmogorov-Smirnov test was carried out comparing the present sample and the average of the other samples. The confidence that the samples are consistent was 95% for the $x_1$ distribution and 58% for the c distribution, both acceptable confidence levels. The discrepancy in the c distribution comes from the reddened tail in the last two bins of Fig. 5. We note that two of the three outliers to the Hubble diagram of Fig. 8, discussed in section 4.5 above, are in this tail. Removing these two outliers from the distribution, the**

**consistency between the c distributions from the Kolmogorov-Smirnov test improves to 90%.**

*5.2 The parameters M, α, and β*

The fitted values of M, α, and β listed in Table 4 are compared to the measurement of these parameters for other samples of supernovae in Table 5 below.

**Table 5**
Comparison of the Parameters M, α, and β*

| Sample | M | α | β |
|---|---|---|---|
| **SNLS** | -19.08±0.03 | 0.138±0.009 | 3.024±0.107 |
| **SDSS** | -19.02±0.03 | 0.145±0.007 | 3.059±0.093 |
| **Union2** | -19.31±0.01 | 0.121±0.007 | 2.510±0.070 |
| **PanSTARRS** | | 0.147±0.010 | 3.13 ±0.12 |
| **LSQ-CSPII** | -19.07±0.03 | 0.13 ±0.05 | 2.23 ±0.30 |

*Notes on Table 5: The Super Nova Legacy Survey SNLS data were taken from Table 10 of Betoule et al (2011) for 118 low z (0.01 < z < 0.10) supernovae and 239 SNLS supernovae. The Sloane Digital Sky Survey SDSS data also comes from Table 10 of Betoule et al (2011) for the same 118 low z and 374 SDSS supernovae. The Union2 data set from Table 10 of Amanullah et al (2010) consists of 166 combined low z sample, 129 SDSS, 102 CfA, 74 ESSENCE, 71 SNLS, and 16 Hubble Space Telescope supernovae. The PanSTARRS data comes from Table 5 of Rest et al (2014) from 197 low z and 113 higher z PanSTARRS supernovae.

There is some spread in the M, α, and β parameters not only with the data sets, but also with the particular analysis of the data. For example there is some variation in these parameters from the SNLS data between the papers of Conley et al (2011), Sullivan et al (2011), and Betoule et al (2010). Sullivan et al (2010) also point out that there is a variation of M with host galaxy properties such as stellar mass. From Table 5 we see that the values obtained in this paper are within the range of the M and α parameters from the other data samples, but is low compared to the others in β, although within one standard deviation of the Union2 data set. This parameter characterizes the correlation between the magnitude of supernovae and their color. This correlation is not well understood at this time. It may be due to extinction of the supernova light between the supernova and Earth or due to intrinsic properties of the supernova itself. The color of the supernova itself may or may not be correlated to its magnitude, as discussed in Mariner et al (2011) and Scolnic et al (2014).

Sullivan et al (2010) used 476 supernovae to find a dependence of M on the galaxy mass, and Scolnic et al (2014) used 518 supernovae to see an effect of the assumptions about the source of the supernova color variation on β. The limited statistics of the present sample of 31 supernovae are not sufficient to contribute to the study of these effects.

*5.3 The Hubble residuals and the supernova intrinsic spread*

The residuals of our data sample in the Hubble diagram of Figure 8 have an rms of 0.156 magnitudes, and the intrinsic spread of the supernovae is $\sigma_{int}$ = 0.14. These numbers are compared to the results from other data samples in Table 6.

**Table 6**
Comparison of the Hubble Residuals and the Intrinsic Spread*

| Sample | rms | $\sigma_{int}$ |
|---|---|---|
| **Low z** | 0.153 | 0.12 |
| **ESSENCE** | 0.20 | 0.13 |
| **SNLS** | 0.156 | 0.08 |
| **SDSS** | 0.143 | 0.11 |
| **Union2** | 0.265 | 0.165 |
| **PanSTARRS** |  | 0.115 |
| **LSQ-CSPII** | 0.156 | 0.14 |

* Same acronyms as in Table 5. The rms values for low z, SDSS and SNLS are taken from Table 4 of Conley et al (2011). The $\sigma_{int}$ values for low z, SDSS

and SNLS are from Table 9 of Betoule et al (2014). The ESSENCE numbers are from Woods-Vasey et al (2007), the numbers for the Union2 data set are from Table 7 of Amanullah et al (2010), and the PanSTARRS number is from Rest et al (2014).

The Hubble diagram residual rms and the intrinsic spread $\sigma_{int}$ are within the range of the other samples in Table 6. Another interesting comparison can be made with the values for the rms and $\sigma_{int}$ that are given in Table 7 of Amanullah et al (2010) for the 16 different supernova data sets, 557 supernovae spanning redshifts from 0.01 to 1.1, that make up the Union2 data set. These numbers are plotted in Figure 9. The values for these two quantities from the present paper are the crosshatched entries on this Figure. They are well within the distribution of these parameters.

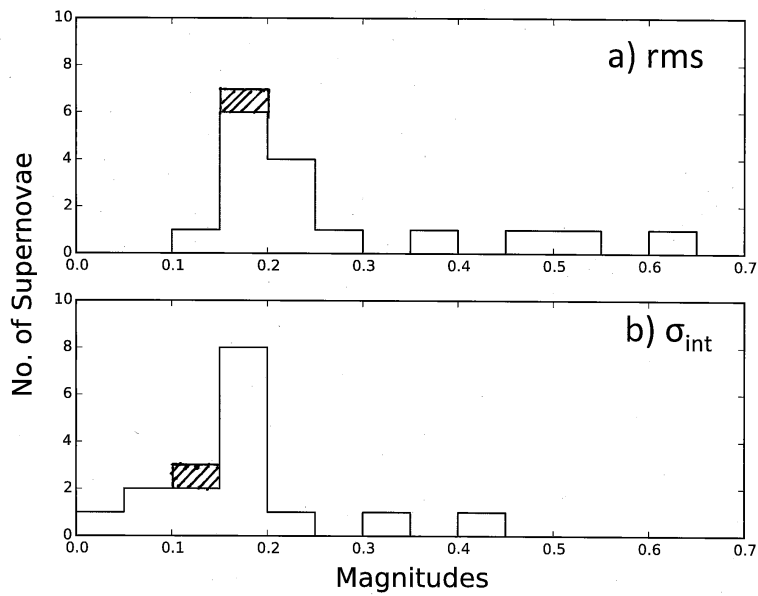

**Figure 9.** Comparison of a) the rms of the residuals of the Hubble diagram and b) the intrinsic spread of the supernova magnitudes $\sigma_{int}$ from the present sample, the crosshatched entry, and the supernova samples listed in Table 7 of the Union 2 data set (Amanullah et al 2010).

*5.4 Future Plans*

The plan of this collaboration is to continue this effort, with possibly other observatories joining, to reach a sample of 300 low redshift SNe Ia in the next few years. The combined total of the presently published nearby supernovae of sufficient quality to use in a cosmological fit is less then 200 (Amanullah et al 2010, Betoule et al 2011, and Rest et al 2014). The other ongoing nearby supernova searches have not published any samples yet. Their expectations are: SNfactory (Aldering et al 2002) expect 300, PTF (Maguire et al 2014) expect 300, and SkyMapper (Keller et al 2007) expect 200 nearby supernovae. Such a combined sample, **even with some expected overlap between the samples,** **should approach roughly a thousand** carefully observed and analyzed supernovae, with some infrared observations by the Carnegie Supernova Project (Philips et al, in preparation). Such a sample will be valuable as a low redshift anchor for high redshift supernova surveys such as SNLS, ESSENCE, DES, LSST and WFIRST. To quantify the effect of such a large sample of nearby supernovae, the WFIRST Science Definition Team (Spergel et al 2015) has estimated the increase in the Figure of Merit (defined as the reciprocal of the area of the $w_a$ v s $w_0$ error ellipse) as a function of the size of the nearby sample. The result, included here as Figure10, shows an increase of a factor of three in the FoM by virtue of a nearby sample of a 1000 supernovae.

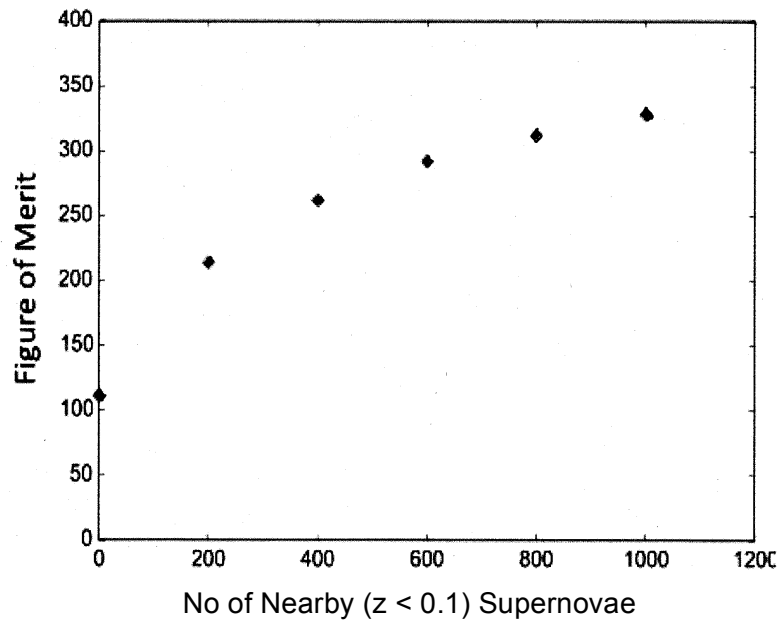

**Figure 10.** The Figure of Merit for the WFIRST supernova survey as a function of the number of nearby supernovae in addition to the 2700 high redshift WFIRST supernovae.

In addition to anchoring the Hubble diagram, which will eventually be systematics limited, a large sample of nearby supernova will be important to sharpen our understanding of the nature of supernovae as cosmological distance indicators (see for example Maguire et al. 2012, Silverman et al. 2013). For example there are indications (Fakhouri et al. 2012) that dividing both the low and the high redshift samples into sub-classes of Type Ia's (comparing twins to twins, so to speak) will reduce systematic uncertainties. A large sample of nearbys will be needed to have sufficient statistics in the individual sub-classes.

ACKNOWLEDGEMENTS

We thank the staffs at Yale University, the Carnegie Observatory, and the La Silla and Las Campanas observatories whose efforts made this supernova survey possible. The Yale group thanks the Office of Science of the US Department of Energy, Grant no DE-FG02-92ER40704 and the Provosts Office at Yale for their support. The CSP acknowledges NSF funding under grants AST-0306969, AST-0908886, AST-0607438, ast-1008343 and support provided by the Danish Agency for Science and Technology and Innovation through a Sapere Aude Level 2 grant.

APPENDIX

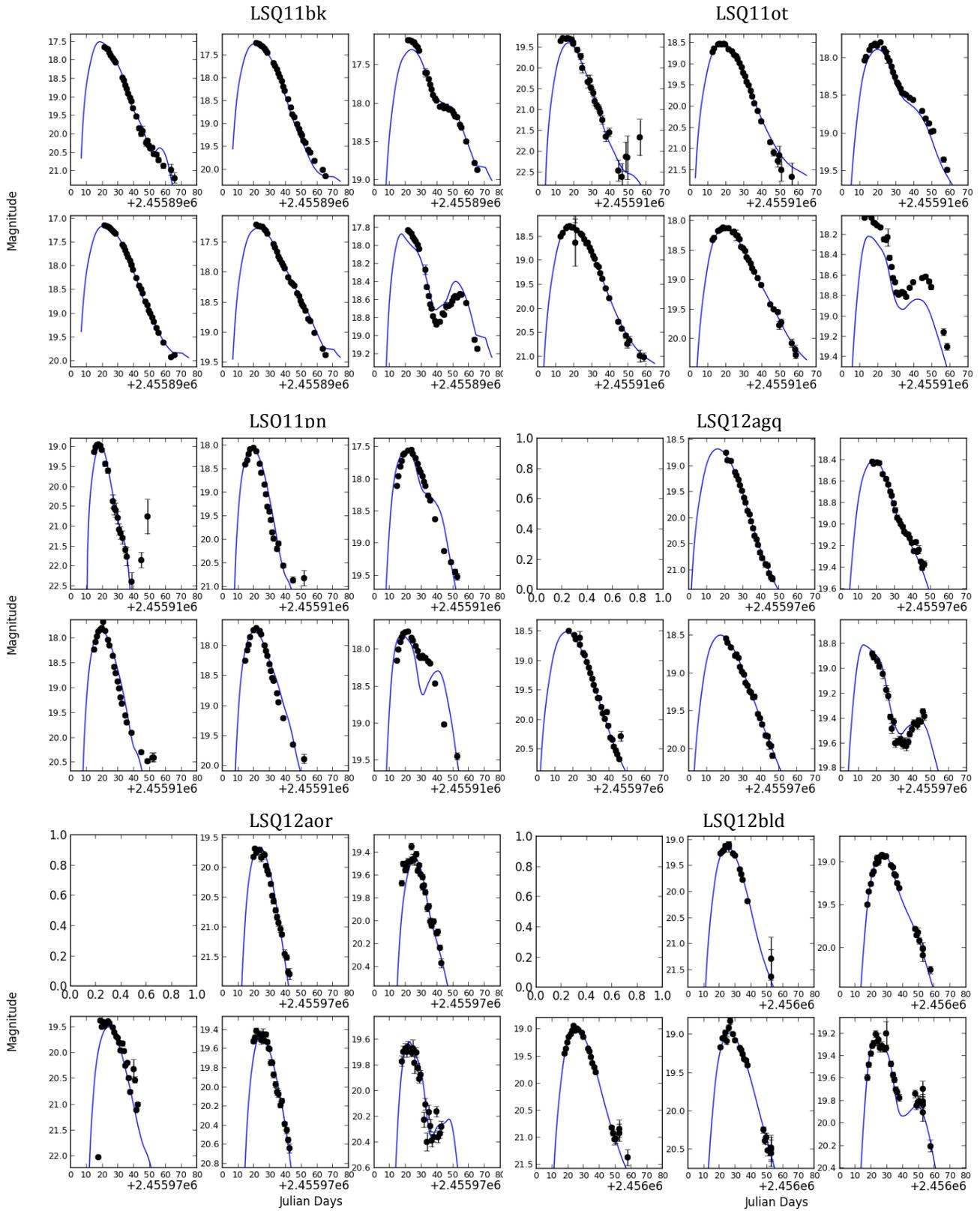

**Figure A1a.** Measured brightness versus Julian Date in filters u, B, r (top row) and g, V, i (bottom row), for supernovae LSQ11bk, LSQ11ot, LSQ11pn, LSQ 11agq, LSQ12aor, and LSQ12bld. In some cases, measurements were not made in all filters. Missing filter observations are left blank.

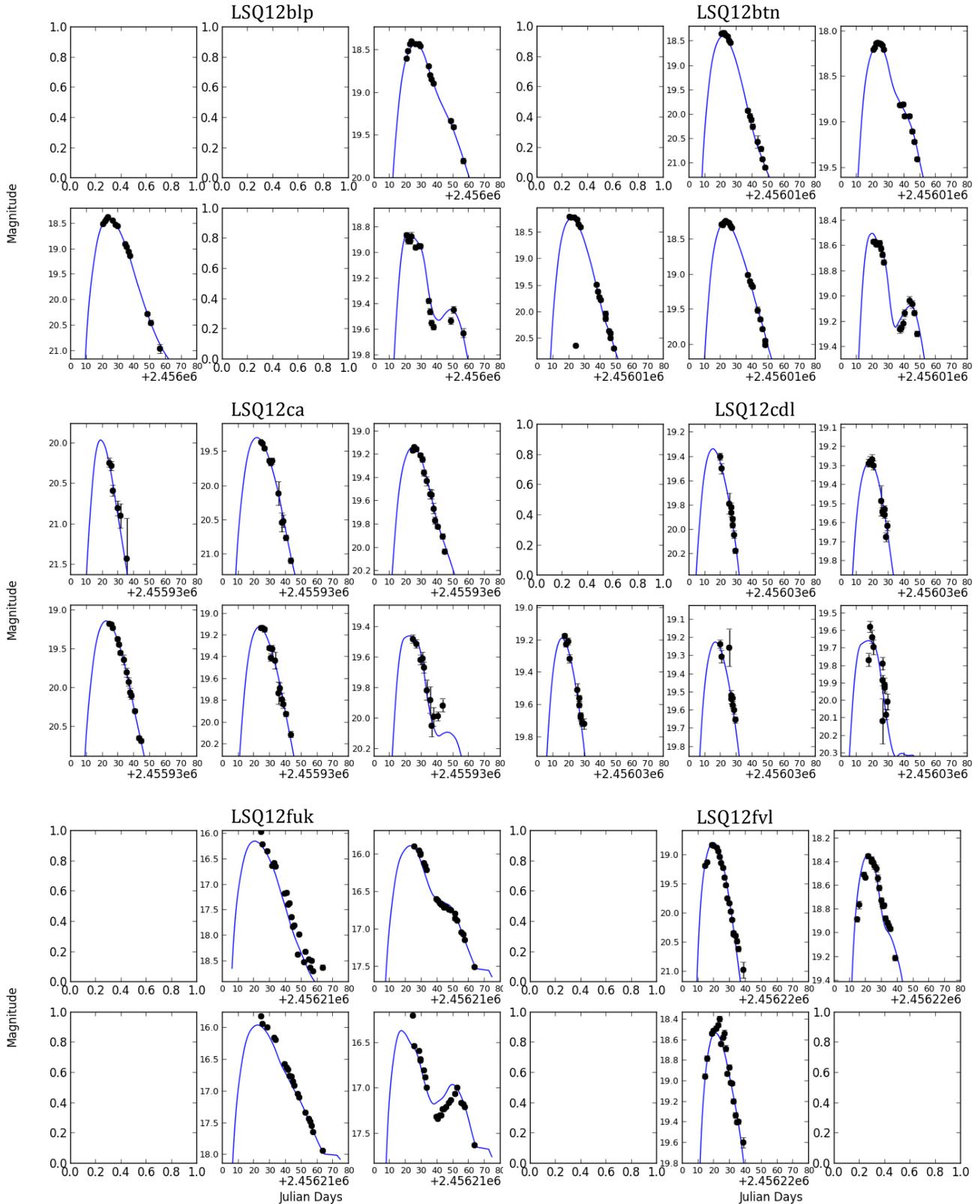

**Figure A1b.** Light curves in filters u, B, r (top row) and g, V, i (bottom row) for supernovae LSQ12blp, LSQ12btn, LSQ12ca, LSQ12cdl, LSQ12fuk, and LSQ12fvl.

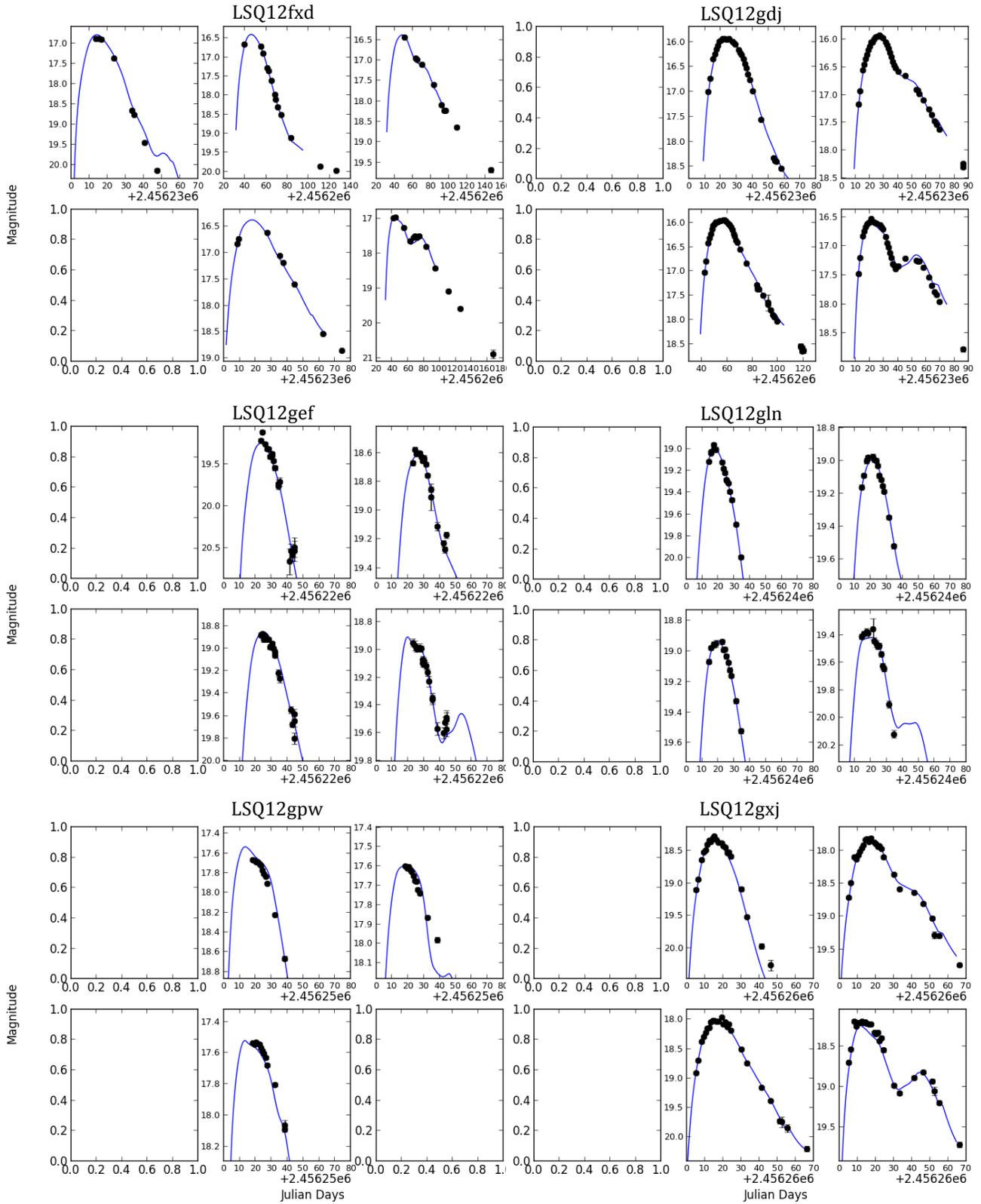

**Figure A1c.** Light curves in filters u, B, r (top row) and g, V, i (bottom row) for supernovae LSQ12fxd, LSQ12gdj, LSQ12gef, LSQ12gln, LSQ12gpw, and LSQ12gxj.

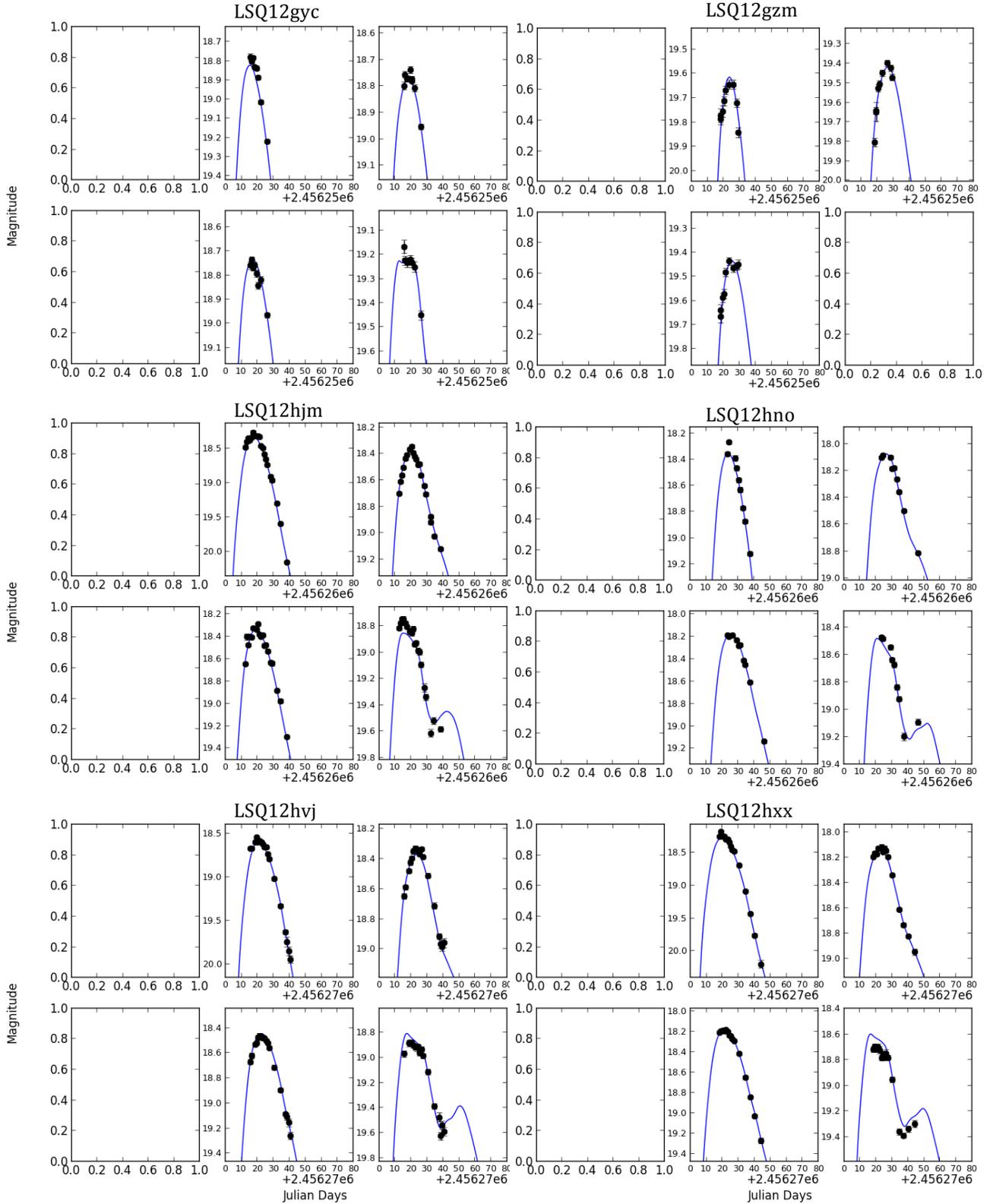

**FigureA1d.** Light curves in filters u, B, r (top row) and g, V, i (bottom row) for supernovae LSQ12gyc, LSQ12gzm, LSQ12hjm, LSQ12hno, LSQ12hvj, and LSQ12hxx.

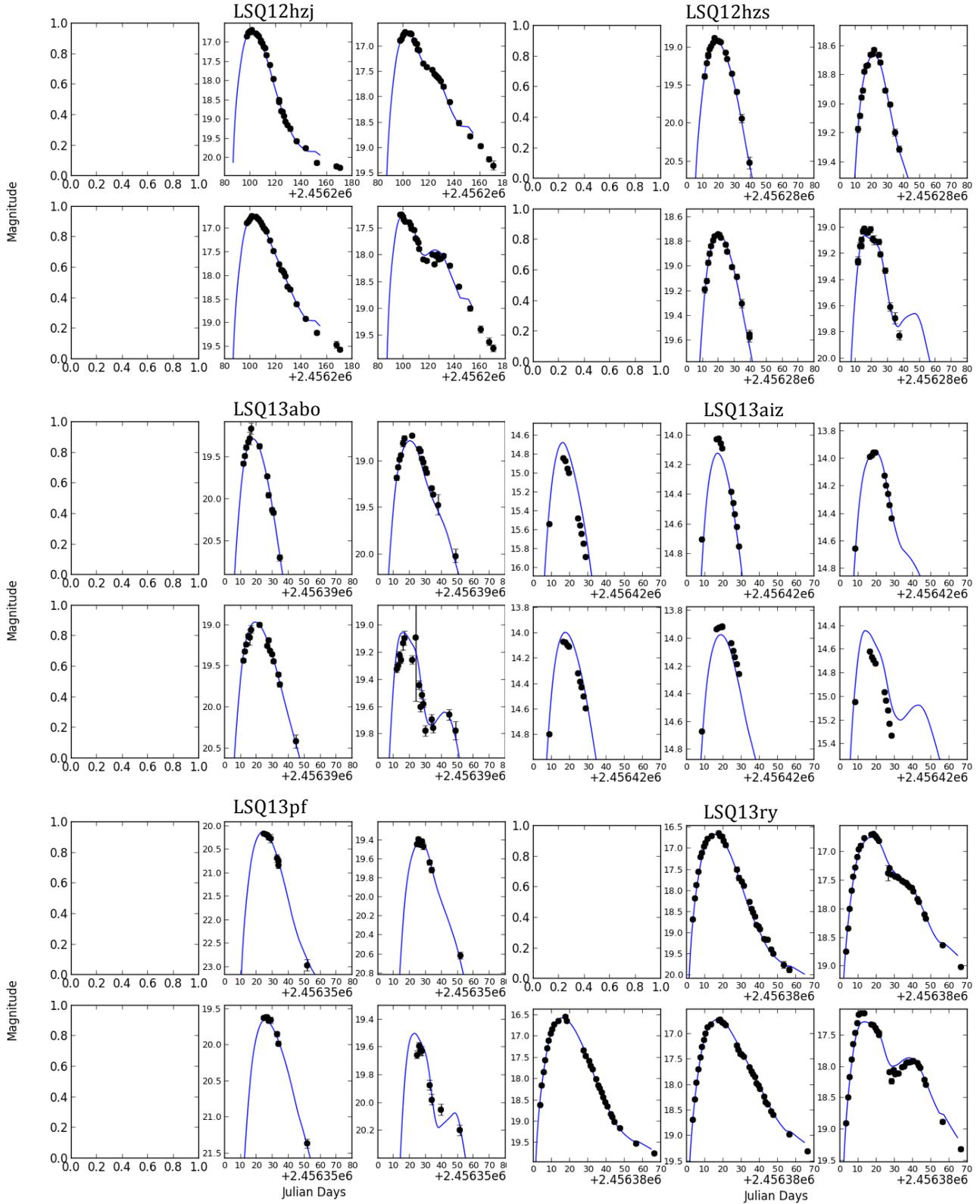

**FigureA1e.** Light curves in filters u, B, r (top row) and g, V, i (bottom row) for supernovae LSQ12hzj, LSQ12hzs, LSQ13abo, LSQ13aiz, LSQ13pf, and LSQ13ry .

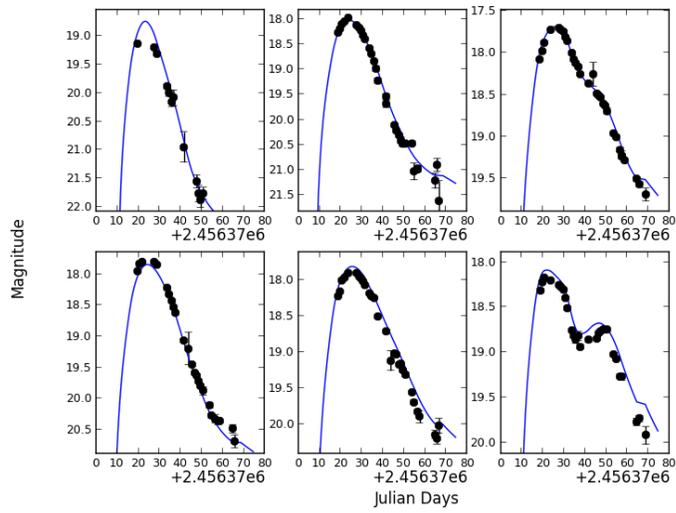

**FigureA1f.** Light curves in filters u, B, r (top row) and g, V, i (bottom row) for supernovae LSQ13vy .